\documentclass[12pt]{article}

\usepackage{amsmath,amssymb,graphicx,cite} 
\usepackage{epsf}
\usepackage{pstricks}

\newcommand{\Tr}{\mathrm{Tr~}}

\newcommand{\beq}{\begin{equation}}
\newcommand{\eeq}{\end{equation}}

\begin{document}
\begin{titlepage}

\vskip.5cm
\begin{center}
{\huge \bf    Odd Decays from Even Anomalies: \\ Gauge Mediation Signatures Without SUSY }

\vskip.2cm
\end{center}

\begin{center}
{\bf {Csaba Cs\'aki}$^a$, {Johannes Heinonen}$^a$, {Jay Hubisz}$^b$, and {Yuri Shirman}$^c$} \\
\end{center}
\vskip 8pt

\begin{center}
$^{a}$ {\it Institute for High Energy Phenomenology\\
Newman Laboratory of Elementary Particle Physics\\
Cornell University, Ithaca, NY 14853, USA } \\

\vspace*{0.3cm}

$^b$ {\it  Argonne National Laboratory, Argonne, IL 60439 USA \\ and \\ Department of Physics, Syracuse University, Syracuse, NY  13244 USA}

\vspace*{0.3cm}

$^c$ {\it Department of Physics, University of California, Irvine, CA 92697 USA}

{\tt  csaki@cornell.edu, jh337@cornell.edu, jhubisz@physics.syr.edu,
yshirman@uci.edu}
\end{center}

\vglue 0.3truecm

\begin{abstract}
\vskip 3pt \noindent We analyze the theory and phenomenology of
anomalous global chiral symmetries in the presence of an extra
dimension.  We propose a simple extension of the Standard Model in
5D whose signatures closely resemble those of supersymmetry with
gauge mediation, and we suggest a novel scalar dark matter
candidate.
\end{abstract}

\end{titlepage}

\newpage

\section{Introduction}
\label{intro}
\setcounter{equation}{0}
\setcounter{footnote}{0}

Anomalies and the interactions they  imply proved crucial in
identifying the ultraviolet physics underlying the chiral
Lagrangian, playing an important role in the formulation of the
dynamical $SU(3)_C$ theory of quarks and
gluons~\cite{BJanom,adleranom,bardeenanom,fuji1,fuji2}.  From the
decay rate of $\pi_0 \rightarrow \gamma \gamma$, for example, one
can infer the number of colors in the UV theory.  This is due to the
fact that, in the $SU(3)_C$ model, anomaly cancellation occurs
non-trivially with the left and right-handed sectors contributing in
equal but opposite non-zero amounts to the anomaly.  In the
effective field theory at low energies, this non-trivial anomaly
cancellation of the UV theory is manifest non-locally in the
$SU(3)_L \times SU(3)_R/SU(3)_V$ theory space of the chiral symmetry
breaking Lagrangian, and emerges as a topological (and thus
quantized) ``Wess-Zumino-Witten" term labelled by a winding number
that corresponds to the number of colors in the UV
theory~\cite{curralg,WZterm,joeanom}.  Additionally, the $U(1)$
problem of QCD, the unexpectedly large masses of the $\eta$ and
$\eta'$ mesons  have been resolved through non-perturbative
instanton contributions through $U(1)$ global
anomalies~\cite{thooftanom}.

As we enter the LHC era, we have identified numerous theories which
may play some role in stabilizing the weak scale.  The most well
studied of these physics scenarios is TeV scale
supersymmetry~\cite{martinprimer}, however, in recent years,
enormous progress has been made on TeV scale extra dimensional
theories and effective field theories such as little Higgs models.
As was the case with the chiral Lagrangian, these theories may be
supplanted at still higher energies by some confining UV dynamics,
and anomalies may again play an important role.  The study of
anomalies in such contexts is in its infancy, but has already
produced some important results for the phenomenology of extensions
of the Standard Model (SM).  To date, most studies have focused on
scenarios where all anomalies vanish in the IR.  In these models,
anomaly cancellation occurs non-locally in an extra
dimension~\cite{nimanom,hillanom}, or, as happens in the chiral
Lagrangian, non-locally in theory space~\cite{hillsquared}.  For
consistency, such theories require a Chern-Simons flux or
Wess-Zumino-Witten term, respectively.  These terms encapsulate the
integrated out UV dynamics through which anomaly cancellation occurs
locally as well as globally.

In this paper, we study the implications  of extra dimensional
classical symmetries which contain  \emph{non-vanishing} anomalies
in the low energy 4D effective theory.  Earlier work on such theories (with some overlapping results) has been performed in~\cite{gripaios}.  The Peccei-Quinn (PQ)
symmetry~\cite{PQ} which has been originally proposed as a solution
to the strong CP problem is a popular and well-motivated example of
such a theory, and thus we consider a $U(1)_\mathrm{PQ}$ extension
of the 5D Universal Extra Dimension model
(UED)~\cite{UED,UEDsplittings,PQinED}. In standard UED, the usual 4D
SM fields are extended so that they all propagate in the bulk of a
compactified extra dimension.  This results in a tower of massive
Kaluza-Klein (KK) partners of each SM field.

We note that this is only one application of the techniques we develop, and that other constructions are possible that may have novel phenomenology.  Examples include warped extra dimensions, or even little Higgs theories, which in certain cases can be related to extra dimensional theories through the language of deconstruction~\cite{deconstruction}.

Even though UED does not explain stability of the weak scale against
radiative corrections, there are several compelling reasons to consider
such theories. In UED there is  remnant of 5D translation invariance
known as KK-parity which stabilizes the lightest KK-mode. Due to
KK-parity tree-level electroweak precision corrections will be
absent (at least from the lightest states), and so these particles
can be quite a bit lighter than the TeV scale. The stability of the
lightest KK mode (LKP) also results in a realistic dark matter
candidate~\cite{UEDDM}. What makes the theory particularly
interesting however is that the UED particle spectrum and collider
phenomenology may be very similar to that of a generic SUSY theory,
and thus UED is a good ``straw-man" to pit against
supersymmetry~\cite{bosonicSUSY}. As in SUSY, the collider
signatures consist of decay chains that contain high $p_T$ jets in
association with large amounts of missing energy. As such, the
models may be difficult to differentiate without resorting to
observables that are sensitive to spin
correlations~\cite{Cambridgespin,PatrickMatt,spincorr,ourspindet},
although techniques are being developed which may be able to
discriminate models in early stages of LHC
running~\cite{lookalikes,maximmoddisc}.

In our study of  this $U(1)_\mathrm{PQ}$ extension of UED (PQ-UED),
we find that anomalies can mediate decays of the KK-odd partners of
the hypercharge gauge boson which is often the lightest KK-odd
particle (LKP), to SM photons and Z's in association with a new
KK-odd scalar field that lives in the 5-component of an
extra-dimensional gauge field.  This $B_5$ is both stable and
neutral, and thus presents as missing energy at colliders.  The
signal event topologies at a hadron collider generically contain
high $p_T$ jets and a pair of neutral SM gauge bosons (either photon
or $Z$).  Final state leptons may also make up a portion of the
event topology, depending on the spectrum of KK-modes.  Such events
are also characteristic of gauge mediated SUSY
breaking~\cite{gaugemed0,gaugemed1,gaugemed2}, where a bino NLSP
decays through a Goldstino coupling to the gravitino plus either a
photon or $Z$.   We thus overturn the lore that such signatures are
a ``smoking gun" for supersymmetry.

This paper is organized as follows. In Section 2, we describe the
basic setup of the PQ-UED model. In Section~\ref{sec:bulkPQ}, we
describe in detail the physics underlying anomalies which persist in
the 4D effective theory.  In particular, we discuss a gauged
$U(1)_\mathrm{PQ}$ symmetry which is broken by boundary conditions
on an $S_1/Z_2$ orbifold.  In Section~\ref{sec:resgts}, we discuss
gauge fixing and the residual gauge transformations, showing that a
massless Goldstone boson results from this choice of boundary
conditions. In~\ref{sec:treeint}, we discuss the tree-level
interactions of the Goldstone boson.  In~\ref{sec:spontbreaking}, we
analyze the physics of additional spontaneous and explicit breaking
of the $U(1)_\mathrm{PQ}$ symmetry, identifying the spectrum and the
wave functions of the physical scalar modes.  In~\ref{sec:5danom},
we discuss quantum mechanical violation of the $U(1)_\mathrm{PQ}$
symmetry, and the interactions of the Goldstone modes that are
generated by the anomalies.  In Section~\ref{sec:pheno}, we study
the  phenomenology of this scenario including collider physics,
discussions about dark matter,  and the existing constraints on the
model (which turn out not to be stringent in the parameter space
that is most interesting from the perspective of collider physics).

\section{Basic Setup}
\label{setup}

The model is in 5D Minkowski space, with the flat distance element:
\begin{equation}
ds^2 = \eta_{\mu\nu} dx^\mu dx^\nu - dz^2,
\end{equation}
where $\eta_{\mu\nu}$ is the metric for 4D Minkowski space.  The
extra dimensional coordinate $z$ is compactified on an $S_1/Z_2$
orbifold, and the $z$-coordinate is taken to range from $z=[0,L]$.
All SM fields are taken to propagate in the bulk, and the Lagrangian
is constructed to obey a discrete $Z_2$ symmetry known as KK-parity,
a remnant of full 5D translation invariance which is broken by the
presence of the branes at $z=0,L$~\cite{UED}.  At the Lagrangian
level, KK-parity forbids bulk Dirac masses for the fermions,
requires that brane localized interactions be identical on the
branes at $z=0,L$, and constrains boundary conditions for bulk
fields to be the same on each brane.  Orbifold boundary conditions
for the fermions and gauge fields are chosen such that the fermion
and gauge boson zero mode spectrum reproduces that of the Standard
Model.  The bulk Higgs sector then gives masses to these modes in
the usual way.

In our setup, we slightly extend UED to incorporate a new bulk gauge
symmetry. This gauge symmetry is chosen to be chiral in the zero
mode spectrum, with the charges matching those of a Peccei-Quinn
global symmetry~\cite{PQ} in Weinberg-Wilczek and DFSZ type axion
models~\cite{WW1,WW2,DFSZ1,DFSZ2}.  In order to do this consistently
we must also have up and down-type Higgs doublets, since the SM with
one Higgs does not have any such symmetry, even at the global level.
In Table~\ref{tab:charges}, we list the charges of the SM fields
under hypercharge and the new gauged PQ symmetry.
\begin{equation}
\label{tab:charges}
\begin{array}{|c|c|c|c|c|c|c|c|}\hline
  \     & H_u & H_d & Q   & \bar{u} & \bar{d} & L         & \bar{e} \\ \hline
Y    & 1/2 & -1/2 & 1/6 & -2/3     &  1/3       &  -1/2  & 1             \\ \hline
\mathrm{PQ}  & 1 & 1  &  -1/2&-1/2   & -1/2      & -1/2   & -1/2         \\ \hline
\end{array}
\end{equation}
Note that a bulk $\mu$ term, $\mu H_u^T (i \tau_2) H_d$, is
forbidden with these charge assignments.   On the boundaries, we fix
the 4D components of the PQ gauge field, $B_M$ to zero:  $B_\mu
|_{z=0,L} = 0$.    In the absence of other symmetry breaking
effects, this leads to a single physical zero mode for the
5-component of the gauge field, $B_5$~\cite{orbifoldBC,goldstoneA5}.
As is normally the case, the remaining KK tower of $B_5$ modes can
be gauged out of the spectrum as they are Goldstone bosons eaten by
the KK tower of massive $B_\mu$ fields.  We discuss this in further
detail in Sections~\ref{sec:resgts} and~\ref{sec:spontbreaking},
where we also take into account bulk breaking of the gauge symmetry
due to the Higgs vacuum expectation values.  Additional explicit
breaking of the $U(1)_\mathrm{PQ}$ symmetry is added in the form of
brane localized $\mu$-terms.  This is done in order to lift a
potential electroweak-scale axion which is ruled out by
experiment~\cite{PDG}.

In this theory, all gauge anomalies (cubic anomalies for gauge
fields with zero modes) vanish as required for consistency.  However
global anomalies (e.g. PQ anomalies quadratic in the SM gauge fields) localized on the
branes at $z=0,L$ persist in the theory~\cite{nimanom}.  These
anomalies lead to couplings of the $B_5$ scalar zero mode to the 5D
field strengths and their duals, $G\tilde{G}$, $W\tilde{W}$ and
$F\tilde{F}$. These couplings allow a decay of the lightest KK-mode in UED, which is often the first KK mode of the hypercharge
gauge boson, down to a photon (or $Z$), and a PQ $B_5$ field.  This
is surprising at first glance, since the $B_5$ has a flat profile,
and is thus naively even under KK-parity.  However, we show in
Section~\ref{sec:treeint} that the zero mode $B_5$ is in fact a
KK-odd field in all of its interactions at both the classical and
quantum levels.

\section{The Gauged Peccei-Quinn Symmetry}
\label{sec:bulkPQ}

In this section, we illustrate the physics  underlying a gauge
symmetry which is broken by boundary conditions at both branes in an
extra dimension constructed on an $S_1/Z_2$ orbifold.  First we
perform gauge fixing, identifying the residual gauge symmetries.
Then we study the interactions of the lowest lying mode, a scalar
field arising from the 5-component of the gauge field, and look at
the implications of additional spontaneous breaking of the gauge
symmetry via a Higgs mechanism. We end with an analysis of anomalies
of this symmetry and the interactions they imply.

\subsection{Residual Gauge Transformations}
\label{sec:resgts} As described in the previous section,  we gauge a
$U(1)_\mathrm{PQ}$ symmetry in the bulk, and break this symmetry via
boundary conditions on the branes at $z=0$ and $z=L$.  In this
section, we analyze this theory, identifying the residual gauge
symmetry after imposing the boundary conditions on the branes, and
adding gauge fixing terms in the bulk which decouple the unphysical
modes.

Requiring preservation of the boundary conditions by the gauge
transformations, $B_M \rightarrow B_M + \partial_M \beta(x,z)$,
gives:
\begin{equation}
\left. B_\mu \right|_{z=0,L} = 0 \quad  \implies \quad  \left. \partial_\mu \beta(x,z) \right|_{z=0,L} = 0.
\label{bcgauge}
\end{equation}
This condition requires that the gauge transformation on the branes
is a constant  function of the 4D coordinates, or is a global
symmetry from the perspective of the 4D theory at $z=0,L$.

We now turn to  gauge fixing the $U(1)_\mathrm{PQ}$ in the bulk.
The 5D Lagrangian for a free $U(1)$ gauge field is given by:
\begin{eqnarray}
&&{\mathcal L}_{U(1)_\mathrm{PQ}} = - \frac{1}{4 g_\mathrm{PQ}^2}   \int dz  B_{MN} B^{MN} \nonumber \\
&=& -  \frac{1}{4 g_\mathrm{PQ}^2}   \int dz  \left[ B_{\mu\nu} B^{\mu\nu} - 2 ( \partial_5 B_\mu)^2 - 2 (\partial_\mu B_5 )^2 + 4 (\partial_5 B^\mu) (\partial_\mu B_5) \right] \nonumber \\
&=& - \frac{1}{4 g_\mathrm{PQ}^2}  \int dz   \left[ B_{\mu\nu}
B^{\mu\nu} - 2 ( \partial_5 B_\mu)^2 - 2 (\partial_\mu B_5 )^2 + 4
(\partial_\mu B^\mu) (\partial_5 B_5) \right]
\nonumber 
-\left. \frac{1}{g_\mathrm{PQ}^2}  B^\mu \partial_\mu B_5
\right|^L_0, \nonumber \\
\end{eqnarray}
where we have rearranged the interaction that  mixes $B_\mu$ and
$B_5$ through integration by parts in the last step.  Note that the
boundary localized term vanishes for the boundary conditions that we
have chosen, $B_\mu |_{L,0} = 0$, so there is no brane localized
mixing between $B_5$ and $B_\mu$.

As we gauge fix, it is convenient to  remove the terms that mix
$B_5$ and $B_\mu$ in the bulk.  This is achieved by adding a gauge
fixing term to the Lagrangian given by~\cite{gaugefixing}:
\begin{equation}
{\mathcal L}_\mathrm{GF} = - \frac{1}{2} \int dz   G^2 \equiv  -\frac{1}{2 g_\mathrm{PQ}^2 \xi_B} \int dz
 \left[ \partial_\mu B^\mu - \xi_B
\partial_5 B_5 \right]^2.
\end{equation}
Note that there is a residual gauge symmetry where the gauge transformation parameter obeys the following equation:
\begin{equation}
\partial_\mu \partial^\mu \beta(x,y) - \xi_B \partial_5^2 \beta(x,y) = 0.
\end{equation}
We choose to go to unitary  gauge, $\xi_B \rightarrow \infty$ where
the eaten $B_5$ modes are projected out of the spectrum.  In this
limit, the solutions are:
\begin{equation}
\beta(x,z) = \beta^+ (x) + \left( \frac{2z - L}{2L} \right) \beta^- (x) \implies \beta_\mathrm{res}(x,z) = \beta^+ + \beta^- \left( \frac{2 z - L}{2L} \right),
\end{equation}
where we have imposed the  boundary conditions in
Eq.~(\ref{bcgauge}) for the gauge transformation in the second step.

Under this residual transformation, the PQ gauge fields transform as:
\begin{equation} \begin{split}
B_\mu & \rightarrow B_\mu \\
B_5 & \rightarrow B_5 + \frac{\beta^-}{L}.
\end{split} \end{equation}
Thus the remaining physical $B_5$ zero mode behaves as a Goldstone
boson, undergoing a constant shift under the KK-odd part of the
residual gauge transformation.  This implies that the choice of
these boundary conditions is equivalent to having spontaneously
broken a global symmetry.  As we will show explicitly in
Section~\ref{sec:5danom}, the effective scale of this symmetry
breaking is given by $f_\mathrm{PQ} = ( g_{5D}^\mathrm{PQ} \sqrt{L}
)^{-1}$.  For the remainder of our analysis, we replace the gauge
coupling with this effective breaking scale using this relation.

Note that the constant transformations $\beta^+$ correspond to a
true (unbroken) PQ global symmetry in terms of the transformation
properties of the light SM fields.  This residual transformation is
unbroken at this stage, and thus the $B_5$ cannot play the role of a
usual axion in resolving the strong CP problem (the $B_5$ is not a
traditional PQ axion).

Before discussing the interactions of the light $B_5$, it is useful
to understand this pattern of symmetry breaking in the language of
deconstruction~\cite{deconstruction}.  This model can be
deconstructed as a chain of $U(1)$ symmetries linked by scalar
fields which each transform under two neighboring $U(1)$ sites.  To
mimic the choice of boundary conditions we have chosen, we only
gauge the internal sites, and the endpoints of the chain are taken
to be global symmetries.  In total, we have $N$ sites, and $N-2$ of
the sites are gauged.  There are $N-1$ scalar fields breaking this
set of symmetries, so there remains one unbroken $U(1)$ symmetry,
corresponding to $\beta^+$ in the continuum theory.  There are $N-1$
Goldstone bosons, and $N-2$ are eaten since $N-2$ of the sites were
gauged.  The remaining physical Goldstone mode corresponds to a non-trivial linear combination of $U(1)$'s and becomes a Wilson line for $B_5$ in the continuum limit.

\subsection{Tree level interactions of the $B_5$ zero mode}
\label{sec:treeint} In this section, we study the interactions  of
the PQ $B_5$ with the KK-modes and SM fields.  In doing so we dispel
the notion that the KK-parity transformation properties of a KK mode
are determined solely by the transformation properties of the wave
function.

This can be seen in a simple way.  First we note that 5D gauge
invariance associates every $\partial_5$ with a $B_5$ and vice versa
through the covariant derivative:
\begin{equation}
D_5 = \partial_5 - i q B_5.
\end{equation}
The form of the 5D flat space metric requires that  any index must
be repeated an even number of times in any single term in the
Lagrangian.\footnote{Except in the case of contraction through the
5D Levi-Civita tensor, however such terms explicitly violate KK
parity as they correspond to a net $U(1)_\text{PQ}$ flux along the
extra dimensional coordinate.}  This is because everything must be
contracted through the metric tensor (or through the vielbeins).
This means that for interactions with an odd number of $B_5$'s,
there must be an odd number of $\partial_5$'s (or a $\gamma^5 \equiv
e^5_a \gamma^a$).  Since both of these pick up a sign under the
transformation $z\rightarrow L-z$, the parity transform of the tower
of $B_5$'s is effectively the opposite of how the wavefunctions
transform. In short, the internal KK parity of the 4D $B_5$ zero mode is
$-$.

As a concrete example, we consider the tree level interactions  with
a 5D fermion.  The interactions arise from the 5D gauge covariant
kinetic term:
\begin{equation}
\label{eq:b5fint}
{\mathcal L}_\mathrm{eff} = \int dz \bar{\Psi} i D_M e^M_a \gamma^a \Psi \supset q\int dz \bar{\Psi} B_5 e^5_a \gamma^a \Psi
\end{equation}
The 5D Dirac fermion can be expanded in solutions of the 4D Dirac equation with masses $m_n$:
\begin{equation}
\Psi = \sum_n \left( \begin{array}{c} g_n(z) \chi_n (x) \\ f_n (z) \bar{\psi}_n (x) \end{array} \right)
\end{equation}
The boundary conditions that produce a $\chi_0$ massless mode are
$f_n (z=0,L) = 0$.   Choosing these boundary conditions, the
solutions for $f_n$ and $g_n$ are given by:
\begin{eqnarray}
&&g_n = A_n \cos \frac{n \pi z}{L} \nonumber \\
&&f_n = - A_n \sin \frac{n \pi z}{L}
\end{eqnarray}
with $A_0 = 1/\sqrt{L}$, and $A_n = \sqrt{2/L}$ for $n \ne 0$.  This
choice reproduces canonically normalized fields in the 4D effective
theory.

We now expand Eq.~(\ref{eq:b5fint}) in KK modes and integrate over $z$, finding
\begin{eqnarray}
\label{eq:b5feyn}
&&{\mathcal L}_\mathrm{eff} = -\frac{1}{f_\mathrm{PQ} L} q \sum_{m,n} c_{nm} B_5(x) \left[ \psi_n \chi_m - \bar{\chi}_m \bar{\psi}_n \right] \nonumber \\
&&c_{nm} = \left\{ \begin{array}{ll} \frac{4}{\pi} \frac{n}{m^2-n^2} & m+n~\mathrm{odd},m \ne 0 \\ \frac{2 \sqrt{2}}{\pi n} &  m+n~\mathrm{odd},m = 0 \\ 0 &m+n~\mathrm{even} \end{array} \right. .
\end{eqnarray}

The $B_5$ is thus a KK-odd field in its interactions with fermions.
The tree-level interactions with scalars  are simpler to calculate,
and the result is similar.  At tree level, the massless $B_5$ is
KK-odd in all of its interactions.

\subsection{Spontaneous Breaking in the Bulk}
\label{sec:spontbreaking}

When the SM Higgs fields obtain vacuum expectation values,  the
$U(1)_\mathrm{PQ}$ symmetry undergoes additional spontaneous
breaking in the bulk.  We show that, in the absence of additional
explicit breaking, the Higgsing along with the choice of boundary
conditions produces two massless modes. One of these is a KK-even
would-be electroweak scale axion that must be lifted, as such a
scalar has interactions that are too strong to remain consistent
with bounds from nuclear and astro-particle physics~\cite{PDG}.  The
other is the KK-odd zero mode whose phenomenology we are most
interested in. Both modes will now be partly contained in $B_5$ and
in the Goldstone field $\pi$ in the bulk Higgs. In this subsection
we first identify these two modes, and then show that an explicit
symmetry breaking term (which is allowed on the boundaries) will
give a mass to both of these states. First we use a simplified
version with a single bulk Higgs, and then show that it is easy to
find the full answer for the two Higgs doublet case relevant for the
bulk U(1)$_{PQ}$ model.

\subsubsection*{The two Goldstone zero modes}

The Lagrangian, before gauge fixing, in our toy model is given by
\begin{equation}
{\mathcal L} = \int dz \left[ \frac{1}{L} | D_M H |^2 - V(H) - \frac{f_\mathrm{PQ}^2 L}{4} B_{MN} B^{MN}Ê\right] -V_\mathrm{bound}  (H_{|_0}) -V_\mathrm{bound}
(H_{|_L}). \label{eq:toymod}
\end{equation}
With the assumption that there are  no brane localized scalar
potential terms, the Higgs develops a z-independent vev profile.
For now, we assume that this is the case, and add brane localized
interactions later, treating them perturbatively in the low-energy
4D effective theory.

First, as in Section~\ref{sec:resgts},  we identify the interactions
which kinetically mix the gauge bosons with the Goldstone bosons, so
that we can remove them with a suitable gauge fixing term.  Taking
$H \equiv \frac{v}{\sqrt{2}} e^{i \pi/v}$, keeping only the
Goldstone fluctuations $\pi$, we have:
\begin{equation}
{\mathcal L}_\mathrm{mix} = -f_\mathrm{PQ}^2  L (\partial_5 B^\mu)
(\partial_\mu B_5) -\frac{1}{L} v \partial_\mu \pi B^\mu
\end{equation}
A gauge fixing term that removes the 4D kinetic mixing is:
\begin{equation}
{\mathcal L}_{\mathrm{GF}} = - \frac{1}{2}  G^2 =
-\frac{f_\mathrm{PQ}^2 L}{2 \xi} \left[ \partial_\mu B^\mu - \xi
\left( \partial_5 B_5 - \frac{1}{f_\mathrm{PQ}^2 L^2} v \pi \right)
\right]^2.
\end{equation}
The residual gauge symmetry obeys the following boundary conditions:
\begin{equation}
\partial_\mu \partial^\mu \beta - \xi  \left( \partial_5^2 \beta - \frac{v^2}{f_\mathrm{PQ}^2 L^2} \beta \right) = 0
\end{equation}
In the $\xi \rightarrow \infty$ limit,  with constant $v$, the
solutions to the equation with appropriate boundary conditions are:
\begin{equation}
\label{eq:ressym} \beta(x,y) = \beta^+ \cosh( \kappa (z - L/2) )   +
\beta^- \sinh( \kappa (z - L/2) ),
\end{equation}
where we have introduced an expansion parameter $\kappa \equiv
v/(f_\mathrm{PQ} L)$. We can find the Goldstone-like zero modes
which are shifting under $\beta$ by carefully analyzing the bulk
EOM's and the BC's, which is performed in detail in Appendix A. The
resulting zero modes can be written in terms of KK even and odd
combinations. In the case where the $B_5$ part has a KK-even
wave-function (but remembering that the interactions are KK-odd) the $B_5$ and $\pi$ zero modes given by
\begin{equation} \begin{split}
B_5^{(0) \mathrm{odd}} & = A'_B \cosh{\kappa (z-L/2)} \zeta_-(x)
\label{eq:GBwf1} \\
\pi^{(0) \mathrm{odd}} & = A'_B \frac{v}{\kappa} \sinh{ \kappa
(z-L/2)}  \zeta_-(x) 
\end{split} \end{equation}
The subtlety about the KK-parity quantum numbers  of the $B_5$ plays
out here, as a single zero mode KK-eigenstate has simultaneous
KK-even and KK-odd wavefunctions (although the interactions are all
consistent, as they must be).

The KK-even modes are given by:
\begin{equation} \begin{split}
B_5^{(0) \mathrm{even}} & = B'_B \sinh{\kappa (z-L/2)} \zeta_+(x) 
\\
\pi^{(0) \mathrm{even}} &=  B'_B \frac{v}{\kappa} \cosh{ \kappa
(z-L/2)}  \zeta_+(x)  \label{eq:GBwf4}
\end{split} \end{equation}
Imposing canonical normalization for the 4D fields  then fixes the
overall coefficients $A'_B$ and $B'_B$. Note that the residual
symmetries in Eq.~(\ref{eq:ressym}) are consistent with the profiles
of these zero modes:  the residual gauge transformations are
shift symmetries for the 4D massless modes, $\zeta_-$ and $\zeta_+$.

\subsubsection*{Explicit brane localized $U(1)_\text{PQ}$ breaking}

We now analyze what happens when we add explicit  symmetry breaking
on the boundaries.  We add PQ breaking $\mu$ terms of the form
$V_\mathrm{bound} = -\frac{\mu}{2} (H^2 + H^{*2})$ on each boundary.
This is allowed, since the symmetry is only global on the endpoints.
Expanding in the Goldstone fluctuations, this leads to brane
localized mass terms for the 5D field $\pi$:
\begin{equation}
V_\mathrm{bound} \Big|_{z=0,L}=   \mu \pi^2 \Big|_{z=0,L}.
\end{equation}
Keeping track of only the (now approximate) zero  modes, this
becomes:
\begin{equation}
V_\mathrm{bound} \Big|_{0,L} =\mu  \left[ A'_B \frac{v}{\kappa} \sinh \kappa (z-L/2) \zeta_- (x) + B'_B \frac{v}{\kappa} \cosh \kappa (z-L/2) \zeta_+ (x) \right]^2 \Big|_{0,L}.
\end{equation}
The effective 4D potential is obtained by summing over the two boundary contributions, which gives:
\begin{equation}
V_\mathrm{eff} =  2 \mu A'^2_B \left( \frac{v}{\kappa} \right)^2 \sinh^2 \frac{\kappa L}{2} \zeta_-^2(x)+ 2 \mu B'^2_B \left( \frac{v}{\kappa} \right)^2 \cosh^2 \frac{\kappa L}{2} \zeta_+^2 (x)
\end{equation}
Expanding in small $\kappa$ and imposing canonical  normalization on
the scalar zero modes in the  4D effective theory takes this to:
\begin{equation}
V_\mathrm{eff} =2 \mu \zeta_+^2 + \frac{1}{2}  \frac{\mu v^2}{f_\mathrm{PQ}^2} \zeta_-^2
\end{equation}
The masses of the KK-even and KK-odd modes are then  $m_+^2 = 4
\mu$, and $m_-^2 = \mu v^2/f_\mathrm{PQ}^2$.  A full numerical
evaluation of the equations of motion, including deformation of the
VEV due to the $\mu$-terms, confirms that these approximations hold
at the level of 2\% for the KK-odd mode, and $<$ 1\% for the KK-even
mode for $\mu$ as large as $(300 ~\mathrm{GeV})^2$.

\subsubsection*{Pseudo-Goldstones in the full 2-Higgs doublet model}

The generalization of this model to the two Higgs  doublet model
(2HDM) of our construction is quite simple.  We first write the two
Higgs doublets keeping only the Goldstone fluctuations along the
$U(1)_\mathrm{PQ}$ flat direction, ignoring the $2$ neutral Higgses,
and the charged Higgs fields.  The Goldstone fluctuation $\pi$ is
the neutral pseudoscalar often referred to as $A_0$ in 2HDMs.
\begin{equation}
H_u = \frac{v_u}{\sqrt{2}} e^{i \pi/V},~~H_d = \frac{v_d}{\sqrt{2}} e^{i \pi/V},~~~\mathrm{with}~~V\equiv \sqrt{v_u^2 + v_d^2}
\end{equation}
In this case, the entire analysis above follows  through the same
way with the replacements
\begin{equation} \begin{split}
v & \rightarrow V =\sqrt{v_u^2+v_d^2} \\
\mathrm{and}~~\mu & \rightarrow \frac{\mu}{2} \sin 2\beta,
\end{split} \end{equation}
where the angle $\beta$ is defined in the usual  way for a 2HDM,
$v_u/v_d \equiv \tan \beta$.  The explicit symmetry breaking terms
in this case are given by
\begin{equation}
{\mathcal L}_\mathrm{mix} = \left. \frac{\mu}{2} H_u^T (i \tau_2) H_d \right|_{z=0,L}
\end{equation}
The final masses are:
\begin{equation} \begin{split}
m_+^2 & = 2 \mu \sin 2 \beta \\
m_-^2 & = \frac{\mu V^2}{2 f_\mathrm{PQ}^2}   \sin 2 \beta .
\end{split} \end{equation}
Taking $\mu_\mathrm{eff} \equiv \frac{\mu \sin 2 \beta}{2}$,  the
numerical expression for the mass of the light pseudo-Goldstone
boson is:
\begin{eqnarray}
m_- &=& (f_\text{PQ} L)^{-1} \left(  \frac{\sqrt{\mu_\mathrm{eff}}}{300~\mathrm{GeV}} \right) \left( L \cdot 10^3~\text{GeV} \right) \cdot 74~\mathrm{GeV}  \nonumber \\
&=& \left( \frac{\sqrt{\mu_\mathrm{eff}}}{300~\mathrm{GeV}} \right) \left(  \frac{10^9~\text{GeV}}{f_\text{PQ}} \right) \cdot 74~\mathrm{keV}.
\end{eqnarray}
For perturbative values of the  coupling $(f_\mathrm{PQ} L)^{-1}$,
and for weak scale $\mu$, the mass of $\zeta_-$ is less than the
mass of any level one KK-mode, whose masses are generally $m^{(1)}
\sim \pi/L$.  So for most choices of parameters, this
pseudo-Goldstone is the LKP.  The reference value of $10^9$~GeV in
the second expression is chosen to match the point at which the
decay length of the NLKP is of order tens of centimeters, as we show
in Section~\ref{sec:pheno}.

\subsection{$U(1)_\mathrm{PQ}$ Anomalies}
\label{sec:5danom}

With the fermion charges given in  Table~\ref{tab:charges}, the
$U(1)_\mathrm{PQ}$ symmetry is anomalous.  However, as we have shown
in Section~\ref{sec:resgts}, the residual symmetry after imposing
Dirichlet boundary conditions on the 4D components of the PQ gauge
field is global on the endpoints of the extra dimension. In this
section we calculate the chiral anomalies in this model, emphasizing
that the chiral anomalies are localized on the branes\cite{nimanom},
where the gauge transformation is global rather than local.   As a
result, the theory is consistent at the quantum level.  However, as
is crucial in our model, the anomalies imply effective interactions
between the $U(1)_\mathrm{PQ}$ $B_5$ and the SM gauge fields.  We
focus on anomalies of the form $U(1)_\mathrm{PQ} \times \mathrm{SM}
\times \mathrm{SM}$, since these lead to the interactions we are
most interested in.

\subsubsection*{An intuitive argument for the localized anomaly terms}

First we present an intuitive argument that suggests the required
form of the localized anomaly terms based on the shift properties of
the action and the Goldstone bosons under the anomalous symmetries.
Later we will give a more rigorous derivation based on the anomalous
transformations of the path integral measure.

Under an anomalous $U(1)_\mathrm{PQ}$ transformation $B_M
\rightarrow B_M +
\partial_M \beta(x,z)$, the action shifts by: \beq \delta
\mathcal{S} = \int d^4x \int_0^L dz~ \beta \partial_M J^M - \int
d^4x \left. \beta  J^5 \right|^L_0 \equiv  \int d^5x ~ \beta
\mathcal{A}, \eeq where $J^M$ is the classically conserved PQ
current, and $\mathcal{A}$ is the anomalous divergence.   The
boundary term vanishes by construction, through the assignment of
the orbifold boundary conditions which produce the chiral spectrum
in Table~\ref{tab:charges}.  The anomaly is itself purely localized
on the branes, and has been calculated in~\cite{nimanom} to be:
\begin{eqnarray}
&\mathcal{A}(x,z) = \frac{1}{2} \left[ \delta(z) + \delta(z-L) \right]  \sum_f q_\mathrm{PQ}^f \left( \frac{q_Y^{f2}}{16 \pi^2} F \cdot \tilde{F}+\frac{\Tr \tau^f_a \tau^f_a}{16 \pi^2} W \cdot \tilde{W}+\frac{\Tr t^f_a t^f_a}{16 \pi^2} G \cdot \tilde{G} \right)& \nonumber \\
&\equiv \frac{1}{2} \left[ \delta(z) + \delta(z-L) \right] {\mathcal
Q}_\mathrm{PQ} (x,z)& \label{eq:nimaanom}
\end{eqnarray}
where $F$, $W$, and $G$  are the hypercharge, $SU(2)_L$, and QCD
field strengths, and $F \cdot \tilde F$ is given by $\frac{1}{2}
\epsilon^{\mu\nu\rho\sigma} F_{\mu \nu} (x,z) F_{\rho \sigma} (x,z)$
(with similar expressions for $W \cdot \tilde{W}$ and $G \cdot
\tilde{G}$).

To reproduce the above shift in the action, the Lagrangian has to
contain a coupling involving the Goldstone bosons, whose shifts will
exactly correspond to the above change in the action. Remembering
that the decomposition of $\beta$ is
\begin{equation}
\beta = \beta^+ \cosh [\kappa (z-\frac{L}{2})]+\beta^- \sinh [\kappa
(z-\frac{L}{2})]
\end{equation}
and the fact that under this shift $B_5\to B_5 +\partial_5 \beta$,
we can identify the shifts of the fields $\zeta_{\pm}$. We find,
that
\begin{equation}
\zeta_{\pm}\to \zeta_{\pm} + v \sqrt{\frac{\sinh \kappa L}{\kappa
L}} \beta^{\pm}.
\end{equation}
Therefore the shift in the action is reproduced if the following
couplings are added to the Lagrangian:
\begin{equation} \begin{split}
{\mathcal L}^\mathrm{eff}_\mathrm{anomaly} & = \frac{1}{2v}  \zeta_{-} \sqrt{\frac{\kappa L}{\sinh \kappa L}} \sinh \frac{\kappa L}{2} \left[ {\mathcal Q}_\mathrm{PQ} (x,L)-{\mathcal Q}_\mathrm{PQ} (x,0) \right] \\
& \qquad +  \frac{1}{2v}  \zeta_{+} \sqrt{\frac{\kappa L}{\sinh
\kappa L}} \cosh \frac{\kappa L}{2} \left[ {\mathcal Q}_\mathrm{PQ}
(x,L)+{\mathcal Q}_\mathrm{PQ} (x,0) \right]. \label{eq:anom1}
\end{split} \end{equation}
To lowest order in the bulk PQ gauge coupling, this becomes:
\begin{equation}
\label{eq:anomL} {\mathcal L}^\mathrm{eff}_\mathrm{anomaly} =
\frac{1}{4 f_\mathrm{PQ}} \zeta_{-}   \left( {\mathcal
Q}_\mathrm{PQ} (x,L)-{\mathcal Q}_\mathrm{PQ} (x,0) \right) +
\frac{1}{2v} \zeta_{+} \left( {\mathcal Q}_\mathrm{PQ}
(x,L)+{\mathcal Q}_\mathrm{PQ} (x,0) \right).
\end{equation}

\subsubsection*{Anomalous interactions from  the path integral measure}

 Above we have seen a
simple argument for the existence of the brane localized
anomalous interactions, motivated by the shifts of the various Goldstone
fields. We now present the full derivation of these terms through the shift in the path integral measure as first identified by Fujikawa~\cite{fuji1,fuji2}. For
this we add two fermions to the single Higgs toy model described by
the effective Lagrangian in Eq.~\ref{eq:toymod}.  These fermions
have $(\pm,\pm)$ and $(\mp,\mp)$ boundary conditions respectively,
such that one fermion has a left handed zero mode, and the other has
a right handed zero mode.  Additionally, they each carry opposite
charge under the $U(1)_\mathrm{PQ}$ symmetry, $q_{L,R} = \pm 1/2$.
The additional terms in the classical effective Lagrangian are:
\begin{equation}
{\mathcal L}_\mathrm{eff}^{\mathrm{fermion}} = \int dz \Big\{
\bar{\Psi}_{L5} i \displaystyle{\empty}{\not} D \Psi_{L5} +
\bar{\Psi}_{R5} i \displaystyle{\empty}{\not} D \Psi_{R5} + \left(
\lambda H \bar{\Psi}_{L5} \Psi_{R5} + \mathrm{h.c.} \right) \Big\}.
\end{equation}
We  now restrict ourselves to the terms in this Lagrangian that
involve the Goldstone bosons $\pi$ and $B_5$:
\begin{eqnarray}
{\mathcal L}_\mathrm{eff}^{\mathrm{fermion}} &\supset& \int dz \Big\{ \bar{\Psi}_{L5} i \left(\partial_5 - i q_L B_5 \right) \gamma^5 \Psi_{L5} + \bar{\Psi}_{R5} i  \left(\partial_5 - i q_R B_5 \right) \gamma^5 \Psi_{R5}  \nonumber \\
&& \qquad \qquad + \left( \frac{\lambda v}{\sqrt{2}} e^{i (q_L-q_R) \frac{\pi}{v}} \bar{\Psi}_{L5} \Psi_{R5} + \mathrm{h.c.} \right) \Big\}.
\end{eqnarray}
We  now perform a redefinition of the fermion fields such that the
new fermion degrees of freedom do not transform under the broken
$U(1)_\mathrm{PQ}$ symmetry.  After this is done, the path integral
measure itself no longer transforms under rotations, and all
interactions of the Goldstone bosons through the anomaly are
manifest.  The redefinition is given by:
\begin{equation}
\label{eq:redef}
\Psi_j = e^{i q_j f (\pi,B_5)} \Psi_j',
\end{equation}
with $f$  transforming as $f \rightarrow f+ \beta(x,z)$, and
$\Psi'_j \rightarrow \Psi'_j$. The most general choice of $f$ that
satisfies this property is a linear combination of a Wilson line and the 5D field $\pi$ from the bulk Higgs:
\begin{equation}
f (\pi,B_5)  = a \left[ \int_{z_0}^z dz' B_5 (x,z') +
\frac{\pi(z_0,x)}{v(z_0)}\right] + (1-a) \frac{\pi(z,x)}{v(z)},
\end{equation}
where $a$ is  an arbitrary c-number.

In terms of  the two physical Goldstone modes, $\zeta_+$ and
$\zeta_-$, the function $f(\pi, B_5)$ is given by:
\begin{equation}
f(\pi, B_5) = \frac{1}{v} \sqrt{\frac{ \kappa  L }{\sinh{\kappa L}}}
\left[ \sinh \kappa (z-L/2) \zeta_- (x) + \cosh \kappa (z-L/2)
\zeta_+ (x) \right]
\end{equation}
It is reassuring  that this result is completely independent of the
two undetermined parameters $z_0$ and $a$.  These parameters are
thus unphysical, and do not affect any interactions after performing
the redefinition.

The redefinition does, however, reorganize other  interactions in
the theory.  The 5D fermion kinetic terms are modified in the
following way at the classical level:
\begin{eqnarray}
\label{eq:kinredef} \bar{\Psi}_j i \displaystyle{\empty}{\not} D
\Psi_j  &=& \bar{\Psi}'_j i D_\mu \gamma^\mu \Psi'_j - q_j
\left(\partial_\mu f (\pi,B_5)\right) \bar{\Psi}'_j \gamma^\mu
\Psi'_j +\bar{\Psi}'_j i \partial_5 \gamma^5 \Psi'_j \nonumber \\
&& \qquad -q_j \left(\partial_5 f(\pi, B_5) - B_5 \right) \bar{\Psi}'_j i
\gamma^5 \Psi'_j .
\end{eqnarray}
Note that this expression is completely gauge invariant under
$U(1)_\mathrm{PQ}$. In addition,  the Goldstone interactions from
the Yukawa term in the Lagrangian become:
\begin{equation}
\frac{\lambda v}{\sqrt{2}} \exp \left[ i \left(q_L-q_R \right) \left( \frac{\pi(z,x)}{v(z)} -  f(\pi,B_5) \right) \right] \bar{\Psi}'_{L5} \Psi'_{R5}
\end{equation}
The argument of this exponential  and the coefficient of the 5D
pseudoscalar current in Eq.~(\ref{eq:kinredef}) are both invariant
under all $U(1)_\mathrm{PQ}$ gauge transformations, and thus these
expressions do not involve either of the physical Goldstone bosons.
This can be verified using the wave functions derived in the
previous section.

It is instructive to compute the  effective 4D currents
corresponding to the broken symmetries associated with the KK-even
and KK-odd pseudo-Goldstone bosons.  At lowest order in the 5D PQ
gauge coupling, the $\zeta_+$ couples diagonally due to wave
function orthogonality, and the current corresponding to this
symmetry is
\begin{equation}
j^\mu_+ = \sum_{j,n} q_j  \bar{\Psi}^{4D}_{j,n} \gamma^\mu
\Psi^{4D}_{j,n}, \ \ \ \ \ \Psi^{4D}_{j,n\ne0} = \left(
\begin{array}{c} \chi_{j,n} (x) \\ \bar{\psi}_{j,n} (x) \end{array}
\right), \ \ \ \ \ \Psi^{4D}_{j,0} = P_j\left( \begin{array}{c}
\chi_{j,0} (x) \\ \bar{\psi}_{j,0} (x) \end{array} \right)
\end{equation}
which can be determined by  reading off the coupling of the
$\zeta_+$ in the 4D effective theory (arising from the second term in~\ref{eq:kinredef}):
\begin{equation}
{\mathcal L}_+ = - \frac{1}{v}\left( \partial_\mu \zeta_+ (x) \right) j_+^\mu,
\end{equation}
where $j$ labels the species  of fermion, and $n$ labels the
KK-level.  The projector, $P_j$ is either $P_+$, or $P_-$, depending
on whether $\Psi_j$ contains a right- or left-handed zero mode.
With the charge assignments we have chosen, from the perspective of
the zero modes, this is an axial-vector current.  The KK-odd current
is more involved:
\begin{eqnarray}
j_-^\mu &=&  \sum_{m,n,j}  q_j c_{mn} \bar{\Psi}^{4D}_{j,m} \gamma^\mu \left[ (m-n)^{-2} + (m+n)^{-2} \gamma^5 \right] \Psi^{4D}_{j,n}. \nonumber \\
c_{mn} &\equiv& \left\{  \begin{array}{ll} 0 & m+n~\mathrm{even} \\
2/\pi^2  & m+n~\mathrm{odd}, m,n \ne 0 \\ \sqrt{2}/\pi^2 & m+n~\mathrm{odd},m\cdot
n=0 \end{array}\right.
\end{eqnarray}
where the coupling is
\begin{equation}
{\mathcal L}_- = - \frac{1}{f_\mathrm{PQ}}\left( \partial_\mu \zeta_- (x) \right) j_-^\mu.
\end{equation}
Note that we have finally  explicitly identified the effective
symmetry breaking scale associated with the $B_5$ Goldstone boson,
justifying our identification $g_{5D} \sqrt{L} \equiv
f_\mathrm{PQ}^{-1}$.

Due to the anomaly, the  redefinition (\ref{eq:redef})  produces a
non-trivial Jacobian in the path integral
measure~\cite{fuji1,fuji2}.  The couplings of the Goldstone bosons
due to the anomaly can then be found by expanding
\begin{equation}
 {\mathcal L}_\mathrm{anomaly}^\mathrm{eff} = \int dz f (\pi,B_5) \mathcal{A}
 \end{equation}
 in terms of the scalar  zero modes. Using again the expression of
 the anomaly from~\cite{nimanom} in (\ref{eq:nimaanom}) we reproduce the expressions (\ref{eq:anom1})-(\ref{eq:anomL}) for the brane localized
 anomalous couplings of the Goldstone bosons.

\subsubsection*{The interactions of $\zeta_-$}

We now turn our focus to the interactions of the KK-odd  Goldstone,
$\zeta_-$, in the effective action Eq.~\ref{eq:anomL}.
Using the KK decomposition of the 5D hypercharge gauge boson (in the
absence of electroweak symmetry breaking), we get
\begin{equation}
F_{\mu \nu}(x,z) = g'_{5D} \sqrt{\frac{1}{L}} F_{\mu \nu}^{(0)}(x) +
g'_{5D} \sum_{n \geq 1} \sqrt{\frac{2}{L}} \cos \left( \frac{n \pi
z}{L} \right) F_{\mu \nu}^{(n)}(x),
\end{equation}
with similar expansions for the $SU(2)_L$ and $SU(3)_C$ field
strengths.   The normalization coefficients are chosen to produce a
canonically normalized 4D effective theory.  This yields
\beq \begin{split} \label{eq:effanomL} 
\mathcal{L}_{B_5 A A}^\mathrm{eff} & =   \frac{1}{16 \pi^2} \frac{1}{f_\mathrm{PQ}}\frac{g'^2_{5D}}{L}  \zeta_- (x) \sum_{m \geq n \geq 0} c_{nm} F^{(n)} \cdot \tilde{F}^{(m)}  \\
   & = \frac{\alpha_1}{4 \pi} \frac{1}{f_\mathrm{PQ}} \zeta_- (x) \sum_{m \geq n \geq 0} c_{nm} F^{(n)} \cdot \tilde{F}^{(m)},
\end{split} \eeq
where $\alpha_1=\frac{g'^2}{4 \pi}$,  $g'=g'_{5D}/\sqrt{L}$ is the
usual 4D effective hypercharge gauge coupling,
$f_\mathrm{PQ}\equiv1/(g^\mathrm{PQ}_{5D} \sqrt{L})$ is the
effective PQ decay constant, and the coefficients $c_{nm}$ are given
by
\beq 
c_{nm} = \left\{ \begin{array}{ll} 0 & \text{$n+m$ even} \\
2 \sum_f q_\mathrm{PQ}^f q_Y^{f2} &  \text{$n+m$ odd, $n,m \geq 1$}
\\ \sqrt{2}  \sum_f q_\mathrm{PQ}^f q_Y^{f2}& \text{$n+m$ odd, $n
\cdot m = 0$.} \end{array}\right.
\eeq

\section{$B_5$ Phenomenology}
\label{sec:pheno}
In this section, we perform a study of the basic phenomenology of this new model.  The collider signatures are quite dramatic: nearly all final state signal events contain high $p_T$ photons or $Z$ bosons along with large amounts of missing energy.  Even more remarkable is that for some ranges of the extra dimensional $U(1)_\text{PQ}$ gauge coupling, the photons or $Z$'s do not generally point back to the original interaction vertex (that is, the  photons or $Z$'s are ``delayed").  Such signatures have long been considered a smoking gun for supersymmetry broken by low scale gauge mediation, and so our analysis suggests that more detailed experimental analyses may be necessary to distinguish supersymmetry from this model.  We calculate the lifetime of the lightest KK-mode and the displacement of the decay vertex from the interaction point.  We assume here that the lightest KK-mode is the level-1 partner of the hypercharge gauge boson.   We also consider the possibility that the $\zeta_-$ Goldstone boson may constitute a large fraction of the observed relic abundance of dark matter, calculating the relic abundance over a range of free parameters in the model.

\subsection{Decays of the NLKP}
We presume that the NLKP is the first  KK-mode of the hypercharge
gauge boson.  This is often the case in UED, since mass splittings
in the level 1 KK sector are achieved at the quantum level through
brane localized kinetic terms.  The small value of $\alpha_1$
implies a smaller contribution to the mass of the level-1
hypercharge gauge boson.\footnote{The level 1-KK mode of the PQ
gauge boson may be lighter, however this mode is \emph{even} under
KK-parity, and additionally has a very small coupling to SM fields.
This particle is thus rarely produced, and does not appear
substantially in the decay products of the KK-modes of SM fields.}
Using the effective Lagrangian in Eq.~(\ref{eq:effanomL}), we
evaluate the matrix element between the level one hypercharge gauge
boson, the $\zeta_-$, and a SM photon or $Z$.  The final
polarization averaged and summed amplitude squared for the decay of
the level-1 KK-mode of the hypercharge gauge boson is given by:
\begin{equation}
\frac{1}{3} \sum_{\mathrm{pol}} \left| i {\mathcal M}_{\gamma,Z}
\right|^2 = \frac{8}{3} \lambda_{\gamma,Z}^2 \left[ \left(p^{(0)}
\cdot p^{(1)}\right)^2 - p^{(0)2}  p^{(1)2} \right] = \frac{2}{3}
\lambda_{\gamma,Z}^2 m^{(1)4} \left[ 1 -
\left(\frac{m^{(0)}}{m^{(1)}} \right)^2 \right]^2
\end{equation}
where $p^{(0)}$ is the momentum of the photon or $Z$, and $\lambda_{\gamma,Z}$ is given by
\begin{equation}
\lambda_{\gamma, Z} = \frac{\alpha_1}{4  \pi}
\frac{1}{f_\mathrm{PQ}} \sqrt{2} \sum_f q_\mathrm{PQ}^f q_Y^{f2}
\cdot (c_w,s_w).
\end{equation}
In the last step, we have evaluated the  products of momenta in the
rest frame of the decaying KK-mode, and we have neglected the mass
of the $B_5$.

For $1/L \gg v$, we can ignore the mass of  the $Z$ boson, and the
partial widths in this limit are given by:
\begin{equation}
\Gamma_{\gamma,Z} \approx  \frac{\alpha^2  }{192 \pi^3 c_w^4
f_\mathrm{PQ}^2} m^{(1)3} \left( \sum_f q_\mathrm{PQ}^f q_Y^{f2}
\right)^2 (c_w^2,s_w^2).
\end{equation}
The sum over charges as can be read in  Table~\ref{tab:charges} is
$\sum_f q^f_\mathrm{PQ} q_Y^{f2} = -5$.   We express the final width
numerically for reference values of the free parameters as:
\begin{equation}
\Gamma_\mathrm{tot} \approx 4.3 \cdot  10^{-7}~\mathrm{eV} \left(
\frac{m^{(1)}}{10^3~\mathrm{GeV}} \right)^3 \left(  \frac{
10^{9}~\mathrm{GeV} }{f_\mathrm{PQ}}\right)^2,
\end{equation}
with branching fractions given by
\begin{equation}
R_\gamma \approx c_w^2~~~~~R_Z \approx s_w^2
\end{equation}
up to terms of order $m_Z^2/m^{(1)2}$.   The total width corresponds
to a lifetime for the NLKP equal to
\begin{equation}
\tau = 1.5 \cdot 10^{-9}~\mathrm{s}  \left(
\frac{10^3~\mathrm{GeV}}{m^{(1)}} \right)^3 \left(
\frac{f_\mathrm{PQ}}{10^{9}~\mathrm{GeV}}  \right)^{2}.
\end{equation}
he NLKP is at the bottom of a decay  chain of exotica produced at a
collider experiment, and the NLKP may travel some measurable
distance before decaying, producing a rather spectacular signature
of high energy photons or $Z$'s which decay to jets or leptons that
do not point back to a central interaction vertex.  The distance
traveled by the NLKP is given by:
\begin{equation}
\Delta x = \gamma v \tau \approx 46~\mathrm{cm}  \left(
\frac{10^3~\mathrm{GeV}}{m^{(1)}} \right)^3 \left(
\frac{f_\mathrm{PQ}}{ 10^{9}~\mathrm{GeV} }\right)^2 \sqrt{ \left(
\frac{E}{m^{(1)} }\right)^2 - 1 }.
\end{equation}
Where $\gamma$ is the relativistic time-dilation  factor, and $v$ is
the velocity.  The typical range for the energy $E$ of the NLKP in a
collider experiment is both model and analysis dependent.  For
larger mass splittings between the different members of the level-1
KK sector, $E$ will typically be larger, as a greater portion of the
parent exotica is converted to kinetic energy.  Also the analyses
performed at collider experiments require specific cuts on the
sample.  For example, an analysis may focus on a trigger sample in
which events are required to contain large amounts of missing
transverse energy.  Such requirements again bias towards larger $E$
for the NLKP, and thus longer decay lengths.

\subsection{$B_5$ Dark Matter}

In the scenario we study, the $B_5$ is most likely  the LKP for all
perturbative choices of the 5D PQ coupling, and is thus a dark
matter candidate when KK-parity is preserved.  In this section, we
discuss the constraints on parameter space based on over-closure
considerations, and the potential of the $B_5$ to make up a
significant fraction of the dark matter relic abundance.  We vary
the scale $f_\text{PQ}$ over a large range, from a standard $\mathcal{O}(1)$
weak coupling to a very high suppression.  An excellent review
that describes the analysis in these different cases can be found
in~\cite{fengdmreview}.

\subsubsection*{The case with weak scale $m_-$}

The gauge coupling may not be very small, in which  case the decays
will be prompt, and the $\zeta_-$ may be a more standard dark matter
candidate, being in thermal equilibrium prior to decoupling.  In
this case, one can evaluate the annihilation cross section, and
follow the usual prescription to evaluate the relic abundance. 
The annihilation to SM particles primarily takes place via
s-channel Higgs exchange.  For our calculation, we assume large $\tan \beta = v_u/v_d$, and that the heavy neutral Higgs is much more massive that the light neutral Higgs:  $m_{H_0} \gg m_{h_0}$.

The thermally averaged non-relativistic annihilation cross section to massive SM gauge fields is given in this limit by:
\begin{equation}
\langle \sigma v \rangle_{W^\pm,Z}= \frac{2 m_-^6}{\pi v_\text{eff}^4} \frac{1}{(4 m_-^2-m_H^2)^2+m_H^2 \Gamma_H^2} \left( 1- \frac{m_V^2}{m_-^2}+\frac{3 m_V^4}{4 m_-^4} \right) \sqrt{1-\frac{m_V^2}{m_-^2}},
\end{equation}
where $v_\text{eff}=246$ GeV is the effective electroweak symmetry breaking scale and $m_V=m_{W,Z}$ is the mass of the massive SM gauge bosons into which the $\zeta_-$ annihilates. 
The annihilation cross section into fermions via the s-channel Higgs in the large $\tan \beta$ and $m_{H_0}\gg m_{h_0}$ limit is given by:
\begin{equation}
\langle \sigma v \rangle_{\bar{f} f}= \frac{m_-^4 m_f^2 }{\pi v_\text{eff}^4} \frac{1}{(4 m_-^2-m_H^2)^2+m_H^2 \Gamma_H^2} \left( 1- \frac{m_f^2}{m_-^2} \right)^{3/2}.
\end{equation}

The annihilation into vectors is rather efficient, even relatively far off of the light Higgs resonance. Thus the preferred band in which the $\zeta_-$ relic abundance saturates the WMAP bound in this mass range is close to the threshold for annihilation into $W$ bosons.  For the annihilation into light fermions, the cross section is suppressed by the fermion mass, and the WMAP window is saturated on the tails of the Higgs resonance.
\begin{figure}[h]
   \centering
   \includegraphics[width=4in]{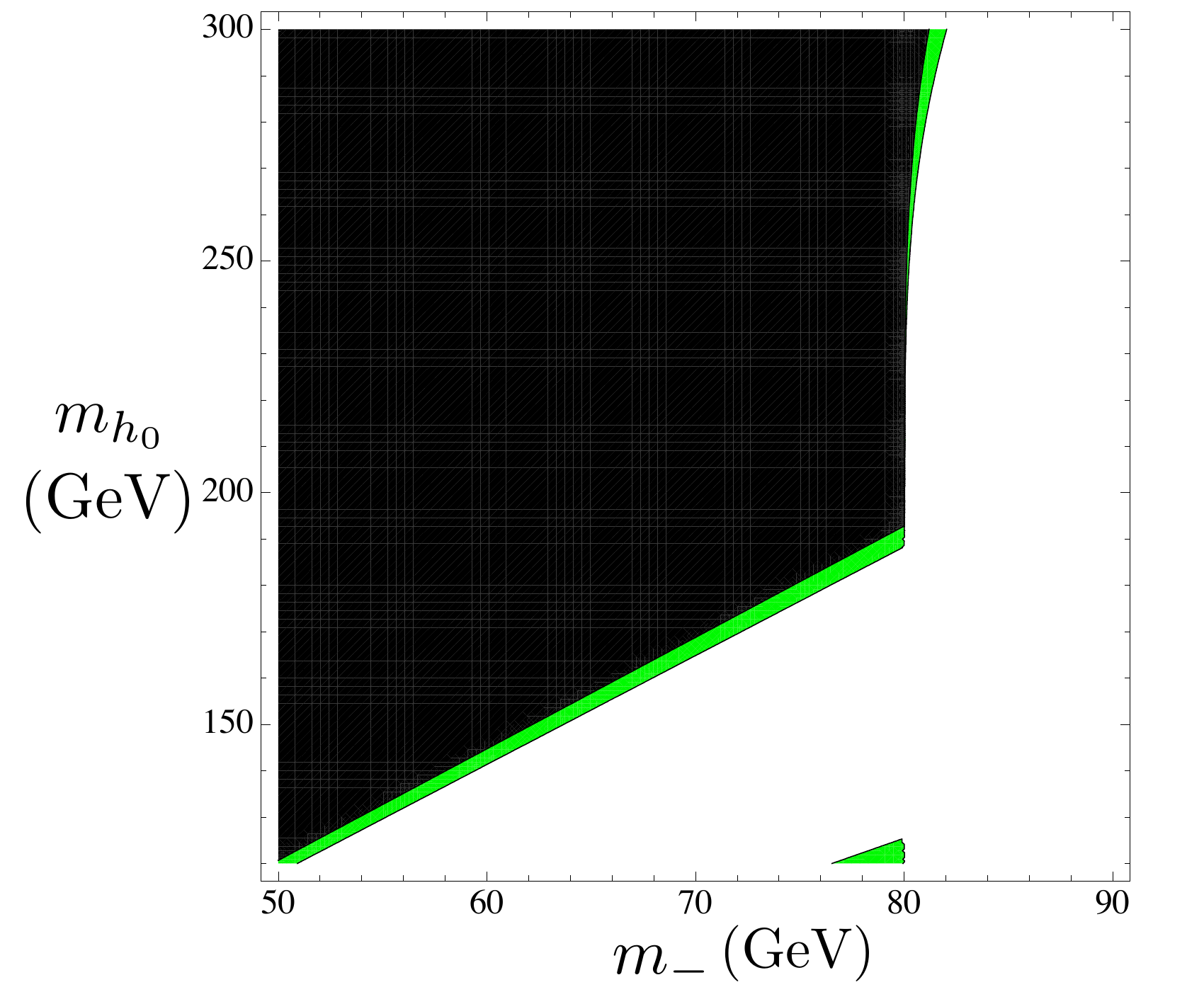}
   \caption{In this Figure, we plot contours of the relic abundance,  $\Omega_\mathrm{dm} h^2$, of the $\zeta_-$ dark matter candidate in the case that the mass of the $\zeta_-$ is near the electroweak scale.   
   The narrow gray band corresponds to the WMAP $2\sigma$ band, where we take the density of non-baryonic dark matter to be $\Omega_\text{nbdm} = .106 \pm .008$~\cite{PDG}.  The white area  corresponds to an under-density of $\zeta_-$ dark matter where it annihilates efficiently, and the dark area corresponds to an over-density.}
\label{fig:dmrelicabundance}
\end{figure}

There are additional channels where the $\zeta_-$ annihilates to photons
or  gluons, however these are essentially two loop diagrams, since
each vertex arises through the anomaly.  These annihilation channels
can thus be ignored.  The results for the relic abundance
calculation are shown in Figure~\ref{fig:dmrelicabundance}.  We plot
contours for when the WMAP result for the relic abundance is
saturated (within the $2\sigma$ band), as well as contours where there is less or more dark matter.

\subsubsection*{The case with low $T_\text{reheat}$, small $(f_\text{PQ} L)^{-1}$}

In the case that the reheating temperature is very low  (on order
the mass of the level 1 KK-modes), and the PQ gauge coupling is
small, the KK-odd Goldstone boson is never in equilibrium with the
thermal bath, and the relic abundance of the $B_5$ in this case
originates primarily from decays of the NLKP.  The final relic
abundance is then given by:
\begin{equation}
\Omega_{B_5} h^2 = \frac{m_{B_5}}{m_\mathrm{NLKP}} \Omega_{\mathrm{NLKP}} h^2.
\end{equation}
The NLKP abundance has been calculated as a  function of mass, and
splittings between KK-modes~\cite{UEDDM,UEDDM2,UEDDM3}. Unless the
the relic abundance of the NLKP is anomalously large, this is
clearly not enough dark matter to saturate the measured relic
abundance.  Of course, in such scenarios, there may be another dark
matter candidate (such as a standard pseudo-scalar axion) which can
make up the remainder.  We note that baryogenesis and leptogenesis
are very problematic in such scenarios, as they must also occur at
this low scale of reheating.

\subsubsection*{The case with larger $T_\text{reheat}$, small $(f_\text{PQ} L)^{-1}$}

In the case where the gauge coupling is small, the universe is
overclosed if the $B_5$ was in thermal equilibrium.   This implies
that some intervening era of inflation must dilute the initial relic
abundance, and that post-reheating, the dark matter never reached
thermal equilibrium with the bath.  The reheat temperature is likely
significantly higher than the mass of the level-1 KK-modes, as is
necessary for generating a baryon asymmetry.  In this case the
situation is considerable more complicated than the previous ones.  The relic abundance in such a scenario
can be found as a function of the reheating temperature and the
couplings to the species which are in equilibrium.  The relic
abundance in this case primarily arises through thermal production
via scattering processes that occur in the bath.

This has been calculated to leading order in the  QCD gauge coupling
for the scenario of a supersymmetric axino DM candidate \cite{axinodm0} in
supersymmetric extensions of the SM~\cite{axinodm}, and the
calculation is quite involved.  In the PQ-UED model, the situation
is even more complicated due to the fact that not only are level-1
KK modes present in the thermal bath, but the entire tower of
KK-modes contributes at a given reheat temperature.  Additionally,
the 5D theory is non-renormalizable, and perturbative unitarity is
lost at energies of order $4 \pi/L$.  The 5D theory must be UV
completed at some relatively low scale, and the characteristics of
this UV completion will likely play a crucial role in the final
relic abundance.  These complications do not by themselves rule out
the potential of the KK-odd Goldstone as a DM candidate in this
region of parameter space, but the calculation is clearly beyond the
scope of this analysis.  We note that it is quite easy to construct
a model that is very similar to that of the MSSM by deconstructing
the extra dimension into a simple 2-site model.  If the symmetry
breaking in this scenario is achieved by a linear sigma model, then
the results would likely be very similar to those in~\cite{axinodm},
with differences arising only from spin statistics in the production
matrix elements, and an extended scalar sector.

In the case of very small $(f_\text{PQ} L)^{-1}$, one might  also worry
about constraints from big-bang nucleosynthesis, or perturbations in
the cosmic microwave background due to the late injection of
electromagnetic energy from NLKP decays.  Neither of these are
relevant for the range of couplings we are most interested in.  BBN
is safe so long as the lifetime of the NLSP is less than 1 second,
the time at which BBN takes place.  This limit on the lifetime, for
weak scale $\mu$, corresponds to a limit on the PQ scale of
$f_\mathrm{PQ} < 10^{14}$~GeV.  The CMB constraints are even more
relaxed, requiring a lifetime of not more than $10^{4-5}$s,
conservatively.  For these large values of the PQ scale, the NLKP
decays far outside of the detector, and does not play a role in
collider physics.

\subsection{Electroweak precision and direct collider constraints}

We estimate the size of shifts in electroweak precision  observables
due to the variation in the vev due to the localized $\mu$ terms.
The terms in the 5D Lagrangian relevant to EWP are:
\begin{equation}
\int dz \frac{g^2 v^2(z)}{8} \left[ W^{(1)2}_\mu + W^{(3)2}_\mu - 2 \frac{g'}{g} W^{(3)}_\mu B^\mu \right]
\end{equation}
We expand the Lagrangian in terms of the KK-modes,  examining the
terms which give mass mixing between the lowest lying modes and the
higher KK-modes.  We treat the vev perturbatively, expanding it as
$v(z) = v_0 + \delta v (z)$.
\begin{equation}
\sum_n \int dz \frac{g^2 v_0 \delta v(z)}{2} \left[ W^{(1)}_{0 \mu}
W^{(1)\mu}_n + W^{(3)}_{0 \mu} W^{(3)\mu}_n - \frac{g'}{g}
W^{(3)}_{0\mu} B^\mu_n \right] \label{eq:ewplag}\end{equation} The
diagrams involving heavy $W$ exchange cancel in  calculating
$\Pi_{11} - \Pi_{33}$, so we need only calculate the diagrams mixing
the heavy $B$ with $W^{(3)}_0$ the last term in
Eq.~(\ref{eq:ewplag}).

We Taylor expand the vev about the midpoint of  the extra dimension,
$\delta v(z) = 1/2 v''(z=L/2) (z-L/2)^2$, and we input the
canonically normalized gauge boson wave functions to find the
relevant overlap integrals for the mixing terms:
\begin{equation}
\frac{g g' v_0 v''_{L/2}}{2 \sqrt{2} L} \int dz (z-L/2)^2 \cos
\frac{n \pi z}{L} =\frac{g g' L^2 v_0 v''_{L/2}}{\sqrt{2} n^2 \pi^2}
\cdot \left\{ \begin{array}{ll} 1 & n~\mathrm{even} \\ 0 &
n~\mathrm{odd} \end{array} \right.
\end{equation}
The diagrams then evaluate to:
\begin{equation}
g^2 \left( \Pi_{11} - \Pi_{33} \right) =  \sum_{n~\mathrm{even}}
\frac{g^2 g'^2 L^6 v_0^2 (v''_{L/2})^2}{2 n^6 \pi^6} =   \frac{g^2
g'^2 L^6 v_0^2 (v''_{L/2})^2}{120,960}
\end{equation}
where  we have used the fact that the masses of the hypercharge
gauge boson KK-modes are approximately given by $m_n = \frac{n
\pi}{L}$. $\Delta \rho$ is then given by:
\begin{equation}
\Delta \rho =  \alpha T = \frac{4}{v_0^2} \left( \Pi_{11} - \Pi_{33}
\right) = \frac{g'^2 L^6 (v''_{L/2})^2}{30,240} \approx 8 \cdot
10^{-9} \left( \frac{ L }{ 1~\mathrm{TeV} } \right)^6 \left(
\frac{\mu}{ 300^2~\mathrm{GeV}^2 }\right)^2,
\end{equation}
well within  current experimental limits.  To understand the overall
scaling with $L$ and $\mu$, remember that $v' |_{0,L} = \mp \mu L v
|_{0,L}$, and thus $v'' \approx \frac{v' |_L - v' |_ 0}{L} \approx 2
\mu v$.

Regarding direct collider constraints,
it is unlikely that the Tevatron experiments searching for GMSB-like scenarios~\cite{D0search,CDFsearch} place any limits on this scenario.  This is due to the fact that there are indirect electroweak precision constraints on the extra dimensional model in addition to the ones calculated above.  These arise from higher dimensional operators in the non-renomalizable 5D theory that are suppressed by the cutoff scale.  Electroweak precision constraints require that this cutoff scale must be at least $5~\text{TeV}$.  These limit the size of the extra dimension to be about $L \lesssim (400~\text{GeV})^{-1}$.  Searches for parity odd quarks in the acoplanar dijet topology at the Tevatron do not yet probe this region of parameter space~\cite{TOQsearch}, and the searches for GMSB like scenarios place even less stringent limits.  The upcoming LHC experiments will have much greater kinematic access to the region which is allowed by electroweak precision constraints.  However, distinction between GMSB scenarios and this extra dimensional model may be difficult given a discovery of an excess of this type of signal.  

\section{Conclusions}
We have performed an  analysis of spontaneously broken anomalous
global symmetries in the context of one universal extra dimension
compactified on an $S_1/Z_2$ orbifold.  A light pseudo-Goldstone
scalar field arises from a 5D gauge symmetry that is broken by
orbifold boundary conditions.  Anomalous couplings to the unbroken
gauge field strengths emerge after performing a 5D field
redefinition that produces a non-trivial Jacobian.  Over a large
range of couplings and explicit symmetry breaking terms, the
resulting effective action permits decays of the lightest level one
SM KK-mode (of the hypercharge gauge boson) to a scalar field
associated with the 5-component of an extra dimensional gauge field
along with either a photon or $Z$-boson.  In particularly
interesting regions of parameter space, the decays occur on detector
sized length scales with sizable displaced vertices.  Such signals
were long thought to be a smoking gun signature of SUSY models in
which the soft masses are generated through gauge mediation, and in
which the NLSP decays to a light gravitino in association with a
photon or $Z$-boson.  We have calculated constraints on this extra
dimensional scenario, finding these to be minimal, and irrelevant
for the range of couplings most interesting from the perspective of
collider phenomenology.  This pseudo-Goldstone scalar field is a
potential dark matter candidate, and it may be possible for it to
saturate the relic abundance observed by WMAP and numerous other
astrophysical experiments.  We have performed a standard relic
abundance calculation for the case in which the extra dimensional
gauge coupling is $\mathcal{O}(1)$.  For small values of the gauge
coupling, the relic abundance calculation is intensive, model
dependent, and depends on unknown details of early cosmology such as
the reheat temperature.  It is unlikely that this region of
parameter space is ruled out by overclosure of the universe, however
the calculation is beyond the scope of this analysis.  BBN and the
CMB spectrum do not place any constraints on the parameter space
most relevant for collider physics.

\section*{Acknowledgments} We thank Jonathan Feng, Gero von Gersdorff, Mark Trodden, Itay Yavin, and Kathryn Zurek for useful
discussions and the KITP at
Santa Barbara for their hospitality while this work was initiated.  We also thank Hsin-Chia Cheng for reading this manuscript prior to submission
and K.C. Kong for pointing out a mistake in the relic density calculation in the first version of this paper.
The work of C.C. is supported in part by the NSF under grant
PHY-0355005 and by a US-Israeli BSF grant. J.He. was supported in
part by the NSF under grant PHY-0355005. J.Hu.~was supported at Argonne National Laboratory under DOE  contract DE-AC02-06CH11357, and by the Syracuse University College of Arts and Sciences.  Y.S. was supported in part by the NSF under grant PHY–0653656.

\appendix

\section*{Appendix}

\section{The Goldstone Wave Functions with a Bulk Higgs VEV}

The full classical equations of motion for $B_5$ and $\pi$ are given
by:
\begin{equation} \begin{split}
\Box \pi - \pi''+ \xi \kappa^2 \pi + \frac{v''}{v} \pi  + ( 1-  \xi) v B'_5  - 2 v' B_5 & = 0  \\
\Box B_5 - \xi B''_5+ \kappa^2 B_5 - (1-\xi) \frac{\kappa^2}{v}
\pi' +(1+ \xi) \frac{\kappa^2}{v^2} v' \pi & = 0
\end{split} \end{equation}
where we have kept the terms containing the derivatives of $v$ for
completeness. After enforcing $B_\mu |_{z=0,L} = 0$, the boundary
conditions for $\pi$ and $B_5$ are given by:
\begin{equation} \begin{split}
\left. \pi' -v B_5-\frac{v'}{v} \pi \pm L \frac{\delta V_\mathrm{bound}}{\delta \pi}  \right|_{z=0,L} & = 0  \\
\left. B'_5 - \frac{\kappa^2}{v} \pi \right|_{z=0,L} & = 0 .
\end{split} \end{equation}

In the cases where $v' = 0$, we can decouple the second order bulk
equations by taking  the first equation, solving for $B'_5$,
\begin{equation}
B'_5 = \frac{1}{v (\xi-1)} \left[ \Box \pi - \pi'' + \xi \kappa^2
\pi \right],
\end{equation}
taking the z-derivative of the second equation, and substituting
using the above formula.  The result is a 4-th order equation for
$\pi$:
\begin{equation}
\pi'''' - 2 \kappa^2 \pi'' + \kappa^4 \pi + m^2 \left\{ (1+1/\xi)
\pi'' + \left[ m^2/\xi - \kappa^2 (1+1/\xi) \right] \pi \right\} = 0
\end{equation}
The same 4-th order equation can be obtained for $B_5$.  Note that
the only dependence on $\xi$ is in the mass terms.  One can
immediately find the physical states (those that don't
depend on $\xi$). For solutions to the second order equation
\begin{equation}
\pi'' + (m^2 - \kappa^2) \pi = 0,
\end{equation}
there is no $\xi$ dependence in the second  half of the equation,
and the bulk eom is also automatically satisfied.  This means that
the remaining two solutions to the full fourth order equation must
be the ones that are eaten/unphysical.

For zero modes, there is trivially no $\xi$ dependence, since $\xi$
appears only in the mass terms.  The most general solutions for the
massless case are:
\begin{equation} \begin{split}
\pi & = A_\pi e^{\kappa z} + B_\pi e^{-\kappa z} + C_\pi z e^{\kappa
z} + D_\pi z e^{-\kappa z}, \\
B_5 & = A_B e^{\kappa z} + B_B e^{-\kappa z} + C_B z e^{\kappa z} +
D_B z e^{-\kappa z}.
\end{split} \end{equation}

We first eliminate 4 of these 8 coefficients  by requiring that the
original second order coupled equations are satisfied.  Satisfying
the boundary conditions further requires that there are no solutions
of the form $z e^{\pm \kappa z}$.  Two undetermined coefficients
remain, implying that there are two physical scalar zero modes in
the spectrum.  The full massless solution is given by:
\begin{equation} \begin{split}
B_5 & = A_B e^{\kappa z} + B_B e^{-\kappa z} \\
\pi & = - \frac{v}{\kappa} \left[ A_B e^{\kappa z} - B_B e^{-\kappa
z} \right].
\end{split} \end{equation}
By rewriting these in KK even and odd combinations we obtain the
final Goldstone wave functions in eqns. \eqref{eq:GBwf1} and \eqref{eq:GBwf4}.


\begin{thebibliography}{99}

\bibitem{adleranom}
  S.~L.~Adler,
  Phys.\ Rev.\  {\bf 177}, 2426 (1969).

\bibitem{BJanom}
  J.~S.~Bell and R.~Jackiw,
  Nuovo Cim.\  A {\bf 60}, 47 (1969).

\bibitem{bardeenanom}
  W.~A.~Bardeen,
  Phys.\ Rev.\  {\bf 184}, 1848 (1969).

\bibitem{fuji1}
  K.~Fujikawa,
  Phys.\ Rev.\ Lett.\  {\bf 42}, 1195 (1979).

\bibitem{fuji2}
  K.~Fujikawa,
  Phys.\ Rev.\  D {\bf 21}, 2848 (1980)
  [Erratum-ibid.\  D {\bf 22}, 1499 (1980)].



\bibitem{curralg}
  E.~Witten,
  Nucl.\ Phys.\  B {\bf 223}, 422 (1983).

\bibitem{WZterm}
  J.~Wess and B.~Zumino,
  Phys.\ Rev.\  {\bf 163}, 1727 (1967).

\bibitem{joeanom}
  O.~Kaymakcalan, S.~Rajeev and J.~Schechter,
  Phys.\ Rev.\  D {\bf 30}, 594 (1984).

\bibitem{thooftanom}
  G.~'t Hooft,
  Phys.\ Rev.\  D {\bf 14}, 3432 (1976)
  [Erratum-ibid.\  D {\bf 18}, 2199 (1978)].

\bibitem{martinprimer}
  For an introduction, see e.g. S.~P.~Martin,
  arXiv:hep-ph/9709356.

\bibitem{nimanom}
  N.~Arkani-Hamed, A.~G.~Cohen and H.~Georgi,
  Phys.\ Lett.\  B {\bf 516}, 395 (2001)
  [arXiv:hep-th/0103135].

\bibitem{hillanom}
  C.~T.~Hill,
  Phys.\ Rev.\  D {\bf 73}, 085001 (2006)
  [arXiv:hep-th/0601154].

\bibitem{hillsquared}
  C.~T.~Hill and R.~J.~Hill,
  Phys.\ Rev.\  D {\bf 76}, 115014 (2007)
  [arXiv:0705.0697 [hep-ph]].

\bibitem{gripaios}
  B.~Gripaios,
  Phys.\ Lett.\  B {\bf 663}, 419 (2008)
  [arXiv:0803.0497 [hep-ph]];
  B.~Gripaios and S.~M.~West,
  Nucl.\ Phys.\  B {\bf 789}, 362 (2008)
  [arXiv:0704.3981 [hep-ph]];
  T.~Flacke, B.~Gripaios, J.~March-Russell and D.~Maybury,
  JHEP {\bf 0701}, 061 (2007)
  [arXiv:hep-ph/0611278].


\bibitem{PQ}
  R.~D.~Peccei and H.~R.~Quinn,
  Phys.\ Rev.\ Lett.\  {\bf 38}, 1440 (1977).



\bibitem{UED}
  T.~Appelquist, H.~C.~Cheng and B.~A.~Dobrescu,
  Phys.\ Rev.\  D {\bf 64}, 035002 (2001)
  [arXiv:hep-ph/0012100].

\bibitem{UEDsplittings}
  H.~C.~Cheng, K.~T.~Matchev and M.~Schmaltz,
  Phys.\ Rev.\  D {\bf 66}, 036005 (2002)
  [arXiv:hep-ph/0204342].

\bibitem{PQinED}
  K.~w.~Choi,
  Phys.\ Rev.\ Lett.\  {\bf 92}, 101602 (2004)
  [arXiv:hep-ph/0308024].


\bibitem{UEDDM}
 G.~Servant and T.~M.~P.~Tait,
  Nucl.\ Phys.\  B {\bf 650}, 391 (2003)
  [arXiv:hep-ph/0206071];
 H.~C.~Cheng, J.~L.~Feng and K.~T.~Matchev,
  Phys.\ Rev.\ Lett.\  {\bf 89}, 211301 (2002)
  [arXiv:hep-ph/0207125].

  \bibitem{bosonicSUSY}
  H.~C.~Cheng, K.~T.~Matchev and M.~Schmaltz,
  Phys.\ Rev.\  D {\bf 66}, 056006 (2002)
  [arXiv:hep-ph/0205314].

\bibitem{Cambridgespin}
 A.~J.~Barr,
  Phys.\ Lett.\  B {\bf 596}, 205 (2004)
  [arXiv:hep-ph/0405052];
J.~M.~Smillie and B.~R.~Webber,
  JHEP {\bf 0510}, 069 (2005)
  [arXiv:hep-ph/0507170].


\bibitem{PatrickMatt}
 P.~Meade and M.~Reece,
  Phys.\ Rev.\  D {\bf 74}, 015010 (2006)
  [arXiv:hep-ph/0601124].



\bibitem{spincorr}
  L.~T.~Wang and I.~Yavin,
  JHEP {\bf 0704}, 032 (2007)
  [arXiv:hep-ph/0605296];
  for a review see L.~T.~Wang and I.~Yavin,
  arXiv:0802.2726 [hep-ph].

\bibitem{ourspindet}
C.~Cs\'aki, J.~Heinonen and M.~Perelstein,
  JHEP {\bf 0710}, 107 (2007)
  [arXiv:0707.0014 [hep-ph]].


\bibitem{lookalikes}
  J.~Hubisz, J.~Lykken, M.~Pierini and M.~Spiropulu,
  Phys.\ Rev.\  D {\bf 78}, 075008 (2008)
  [arXiv:0805.2398 [hep-ph]].

\bibitem{maximmoddisc}
  G.~Hallenbeck, M.~Perelstein, C.~Spethmann, J.~Thom and J.~Vaughan,
  arXiv:0812.3135 [hep-ph].


\bibitem{deconstruction}
  N.~Arkani-Hamed, A.~G.~Cohen and H.~Georgi,
  Phys.\ Rev.\ Lett.\  {\bf 86}, 4757 (2001)
  [arXiv:hep-th/0104005];
  C.~T.~Hill, S.~Pokorski and J.~Wang,
  Phys.\ Rev.\  D {\bf 64}, 105005 (2001)
  [arXiv:hep-th/0104035].


\bibitem{gaugemed0}
  M.~Dine and W.~Fischler,
  Phys.\ Lett.\  B {\bf 110}, 227 (1982);
  M.~Dine and W.~Fischler,
  Nucl.\ Phys.\  B {\bf 204}, 346 (1982);
  L.~Alvarez-Gaume, M.~Claudson and M.~B.~Wise,
  Nucl.\ Phys.\  B {\bf 207}, 96 (1982);
  S.~Dimopoulos and S.~Raby,
  Nucl.\ Phys.\  B {\bf 219}, 479 (1983).

\bibitem{gaugemed1}
  M.~Dine, A.~E.~Nelson and Y.~Shirman,
  Phys.\ Rev.\  D {\bf 51}, 1362 (1995)
  [arXiv:hep-ph/9408384];
  M.~Dine, A.~E.~Nelson, Y.~Nir and Y.~Shirman,
  Phys.\ Rev.\  D {\bf 53}, 2658 (1996)
  [arXiv:hep-ph/9507378].

\bibitem{gaugemed2}
  G.~F.~Giudice and R.~Rattazzi,
  Phys.\ Rept.\  {\bf 322}, 419 (1999)
  [arXiv:hep-ph/9801271].

\bibitem{WW1}
  S.~Weinberg,
  Phys.\ Rev.\ Lett.\  {\bf 40}, 223 (1978).

\bibitem{WW2}
  F.~Wilczek,
  Phys.\ Rev.\ Lett.\  {\bf 40}, 279 (1978).

\bibitem{DFSZ1}
  M.~Dine, W.~Fischler and M.~Srednicki,
  Phys.\ Lett.\  B {\bf 104}, 199 (1981).

\bibitem{DFSZ2}
  A.~R.~Zhitnitsky,
  Sov.\ J.\ Nucl.\ Phys.\  {\bf 31}, 260 (1980)
  [Yad.\ Fiz.\  {\bf 31}, 497 (1980)].

\bibitem{orbifoldBC}
  Y.~Kawamura,
  Prog.\ Theor.\ Phys.\  {\bf 105}, 999 (2001)
  [arXiv:hep-ph/0012125].

\bibitem{goldstoneA5}
  R.~Contino, Y.~Nomura and A.~Pomarol,
  Nucl.\ Phys.\  B {\bf 671}, 148 (2003)
  [arXiv:hep-ph/0306259].

\bibitem{PDG}
  C.~Amsler {\it et al.}  [Particle Data Group],
  Phys.\ Lett.\  B {\bf 667}, 1 (2008).

\bibitem{gaugefixing}
L.~Randall and M.~D.~Schwartz,
  JHEP {\bf 0111}, 003 (2001)
  [arXiv:hep-th/0108114];
A.~Muck, A.~Pilaftsis and R.~Ruckl,
  Phys.\ Rev.\  D {\bf 65}, 085037 (2002)
  [arXiv:hep-ph/0110391];
 G.~Cacciapaglia, C.~Cs\'aki, C.~Grojean, M.~Reece and J.~Terning,
  Phys.\ Rev.\  D {\bf 72}, 095018 (2005)
  [arXiv:hep-ph/0505001].

\bibitem{thermalgrav}
  M.~Bolz, A.~Brandenburg and W.~Buchmuller,
  Nucl.\ Phys.\  B {\bf 606}, 518 (2001)
  [Erratum-ibid.\  B {\bf 790}, 336 (2008)]
  [arXiv:hep-ph/0012052].

\bibitem{fengdmreview}
  J.~L.~Feng,
  [arXiv:hep-ph/0405215].


\bibitem{UEDDM2}
  H.~C.~Cheng, J.~L.~Feng and K.~T.~Matchev,
  Phys.\ Rev.\ Lett.\  {\bf 89}, 211301 (2002)
  [arXiv:hep-ph/0207125].

\bibitem{UEDDM3}
  K.~Kong and K.~T.~Matchev,
  JHEP {\bf 0601}, 038 (2006)
  [arXiv:hep-ph/0509119].

\bibitem{axinodm0}
  T.~Goto and M.~Yamaguchi,
  Phys.\ Lett.\  B {\bf 276}, 103 (1992).

\bibitem{axinodm}
  A.~Brandenburg and F.~D.~Steffen,
  JCAP {\bf 0408}, 008 (2004)
  [arXiv:hep-ph/0405158].

\bibitem{D0search}
  V.~M.~Abazov {\it et al.}  [D0 Collaboration],
  Phys.\ Rev.\ Lett.\  {\bf 101}, 111802 (2008)
  [arXiv:0806.2223 [hep-ex]].


\bibitem{CDFsearch}
  T.~Aaltonen {\it et al.}  [CDF Collaboration],
  Phys.\ Rev.\  D {\bf 78}, 032015 (2008)
  [arXiv:0804.1043 [hep-ex]].

\bibitem{TOQsearch}
  V.~M.~Abazov {\it et al.}  [D0 Collaboration],
  Phys.\ Lett.\  B {\bf 668}, 357 (2008)
  [arXiv:0808.0446 [hep-ex]];
  M.~S.~Carena, J.~Hubisz, M.~Perelstein and P.~Verdier,
  Phys.\ Rev.\  D {\bf 75}, 091701 (2007)
  [arXiv:hep-ph/0610156].

\end{thebibliography}
\end{document}